\documentclass[twocolumn,showpacs,preprintnumbers,amsmath,amssymb]{revtex4}

\usepackage{graphicx}
\usepackage{dcolumn}
\usepackage{bm}
\begin{document}

\preprint{}

\title{LHC would-be $\gamma\,\gamma$ excess as a non-perturbative effect of the electro-weak interaction}

\author{Boris A. Arbuzov }

\affiliation{Skobeltsyn Institute of Nuclear Physics, Lomonosov Moscow State University\\ Leninskie gory 1, 119991 Moscow, Russia}%
\email{arbuzov@theory.sinp.msu.ru}
\author{Ivan V. Zaitsev}
\affiliation{Skobeltsyn Institute of Nuclear Physics, Lomonosov 
Moscow State University\\ Leninskie gory 1,119991 Moscow, Russia}
\date{\today}
\begin{abstract}
The recently reported would-be excess at $125\, GeV$ in  invariant mass distribution of $\gamma\, \gamma$ and of $l^+\,l^+\,l^-\,l^-$ obtained in the course of the Higgs boson search at LHC is tentatively interpreted as a scalar bound state of two $W$.  Non-perturbative effects of EW interaction obtained by application of Bogoliubov compensation approach lead to such bound state due to existence of anomalous three-boson gauge-invariant effective interaction. The application of this scheme gives satisfactory agreement with existing data without any adjusting parameter but the bound state mass $125\,GeV$, while $\sigma\,BR$ for $\gamma\,\gamma$ resonance is predicted to be twice more as the value for the SM Higgs. Decay channel $\gamma\,l^+\,l^-$ may serve as a decisive check of the interpretation.
\end{abstract}

\pacs{12.15.-y; 12.15.Ji; 14.70.Fm; 14.80.Ec}
\keywords{anomalous three-boson
 interaction; the Higgs boson search; W-hadrons}
\maketitle

\section{Strong effective tree-boson interaction}

Recent LHC results on searches for Higgs~\cite{LHC1,LHC2} already induce active discussion. Hints on existence of a state with mass around $125\,GeV$, which manifest itself in decays to $\gamma\,\gamma$ and $l^+l^+l^-l^-$, are interpreted not only in terms of SM Higgs, but also in different variants extensions of the SM: fermiophobic Higgs~\cite{fph3}, two Higgs doublet models~\cite{2doubl} {\it etc}. In any case data being presented in~\cite{LHC1,LHC2} allow discussion of different options the more so, as agreement of the data with SM predictions is not very convincing.

In the present work we would discuss an interpretation of the would-be LHC $125\,GeV$ bump in terms of non-perturbative effects of the electro-weak interaction. For the purpose we rely on an approach induced by N.N. Bogoliubov compensation principle~\cite{Bog1,Bog2}.
In works~\cite{BAA04} - \cite{AZ11}, this approach
was applied to studies of a spontaneous generation of effective non-local interactions in renormalizable gauge theories.
In particular, papers~\cite{BAA09,AZ11}  deal with an application of the approach to the electro-weak interaction and a possibility of spontaneous generation of effective anomalous three-boson interaction of the form
\begin{eqnarray}
& &-\,\frac{G}{3!}\,F\,\epsilon_{abc}\,W_{\mu\nu}^a\,W_{\nu\rho}^b\,W_{\rho\mu}^c\,;
\nonumber\\
& &W^3_{\mu \nu}\,=\,\cos\theta_W\,Z_{\mu \nu}\,+\,\sin\theta_W\,A_{\mu \nu}\,;\label{FFF}\\
& &W_{\mu\nu}^a\,=\,
\partial_\mu W_\nu^a - \partial_\nu W_\mu^a\,+g\,\epsilon_{abc}W_\mu^b W_\nu^c\,.\nonumber
\end{eqnarray}
with uniquely defined form-factor $F(p_i)$, which guarantees effective interaction~(\ref{FFF}) acting in a limited region of the momentum space. It was done of course in the framework of an approximate scheme, which accuracy was estimated to be $\simeq 10\%$~\cite{BAA04}. Would-be existence of effective interaction~(\ref{FFF}) leads to important non-perturbative effects in the electro-weak interaction. It is
usually called anomalous three-boson interaction and it is considered for long time on phenomenological grounds~\cite{Hag1,Hag2}. Note, that the first attempt to obtain the anomalous three-boson interaction in the framework of Bogoliubov approach was done in work~\cite{Arb92}. Our interaction constant $G$ is connected with
conventional definitions in the following way
\begin{equation}
G\,=\,-\,\frac{g\,\lambda}{M_W^2}\,;\label{Glam}
\end{equation}
where $g \simeq 0.65$ is the electro-weak coupling.
The current limitations for parameter $\lambda$ read~\cite{EW}
\begin{eqnarray}
& &\lambda\, =\, -\,0.016^{+0.021}_{-0.023}\,;\nonumber\\
& &  -\,0.059< \lambda < 0.026\,
(95\%\,C.L.)\,.\label{lambda1}
\end{eqnarray}

Interaction~(\ref{FFF}) increases with increasing momenta $p$. For estimation of an effective dimensionless coupling we choose symmetric momenta (p\,,q\,,k) in vertex corresponding to the interaction
\begin{eqnarray}
& &(2\pi)^4\,G\,\,\epsilon_{abc}\,(g_{\mu\nu} (q_\rho pk - p_\rho qk)+ \nonumber\\
& &g_{\nu\rho}
(k_\mu pq - q_\mu pk)+g_{\rho\mu} (p_\nu qk - k_\nu pq)+\label{vertex}\\
& &+\,q_\mu k_\nu p_\rho - k_\mu p_\nu q_\rho)\,F(p,q,k)\,
\delta(p+q+k)\,+...;\nonumber
\end{eqnarray}
where
$p,\mu, a;\;q,\nu, b;\;k,\rho, c$ are respectfully incoming momenta,
Lorentz indices and weak isotopic indices
of $W$-bosons. We mean also that there are present four-boson, five-boson and
six-boson vertices according to expression for $W_{\mu\nu}^a$
(\ref{FFF}). In what follows we shall use four boson vertex,
which corresponds to the following interaction
\begin{equation}
\Delta L\,=\frac{g\, G}{2}\,\epsilon_{abc}\,\epsilon_{aed}\,W_\mu^e\,W_\nu^d\,W_{\nu \rho}^b\,W_{\rho \mu}^c\,.
\label{FFAA}
\end{equation}
Explicit expression for the corresponding vertex is presented in work~\cite{BAA09}.
Form-factor $F(p,q,k)$ is obtained in work~\cite{AZ11} using the following approximate dependence on the three variables
\begin{eqnarray}
& &F(p,q,k)\,=\,F\Bigl(\frac{p^2+q^2+k^2}{2}\Bigr);\label{FP}\\ 
& &F(p,q,k)|_{k=0}\,=\,F(p^2)\,.\nonumber
\end{eqnarray}
Symmetric condition means
\begin{equation}
pq\,=\,pk\,=\,qk\,=\,\frac{p^2}{2}\,=\,\frac{q^2}{2}\,=\,\frac{k^2}{2}\,=\,\frac{x}{2};
\label{symm}
\end{equation}
Interaction~(\ref{FFF}) increases with increasing momenta $p$ and
corresponds to effective dimensionless coupling being of the following order of magnitude
\begin{equation}
g_{eff}\,=\,\frac{|g\,\lambda|\,p^2}{2 M_W^2}\,F\Bigl(\frac{3\,p^2}{2}\Bigr)\,.\label{geff}
\end{equation}
Form-factor $F(x)$ in work~\cite{AZ11} is expressed in terms of the Meijer functions~\cite{BE}
\begin{eqnarray}
& &F(z)\,=\,\frac{1}{2}\,G_{15}^{31}\Bigl( z\,|^0_{1,\,1/2,\,0,\,-1/2,\,-1}
\Bigr) -\nonumber\\
& &\frac{85\,g_0 \sqrt{2}}{128\,\pi}\,G_{15}^{31}\Bigl( z\,|^{1/2}_{1,\,1/2,
\,1/2,\,-1/2,\,-1}\Bigr)\,+\nonumber\\
& &C_1\,G_{04}^{10}\Bigl( z\,|1/2,\,1,\,-1/2,\,-1\Bigr)\,+
\label{solutiong}\\
& &C_2\,G_{04}^{10}\Bigl( z\,|1,\,1/2,\,-1/2,\,-1\Bigr)\,.
\quad z\,=\,\frac{G^2\,x^2}{512\,\pi^2}\,.\nonumber
\end{eqnarray}
\begin{equation}
g_0\,=\,0.6037 ;\;
C_1\,=\,-\,0.0351 ; \; C_2\,=\,-\,0.0511\,;\label{gY}
\end{equation}
where $g_0$ is value of the electro-weak running coupling at momentum
$p_0$ corresponding to value of variable z
\begin{equation}
 z_0\,=\,9.6175\,.\label{z0}
\end{equation}
Thus running $g_{eff}$ in dependence on variable $t\,=\,G\,p^2$
is the following
\begin{equation}
g_{eff}(t)\,=\,\frac{t}{2}\,F\Bigl(\frac{9\,t^2}{2048\,\pi^2}\Bigr)\,;\quad t\,=\,G\,p^2\,.\label{gefft}
\end{equation}
Behavior of $g_{eff}(t)$ is presented at Fig.1. We see that for $t \simeq 22$ the coupling reaches maximal value $g_{eff} \,=\, 3.63$, that is corresponding effective $\alpha$ is the following
\begin{equation}
\alpha_{eff} = \frac{g_{eff}^2}{4\,\pi}\,=\,1.049\,.\label{aleff}
\end{equation}
Thus for sufficiently large momentum interaction~(\ref{FFF})  becomes strong
and may lead to physical consequences analogous to that of the usual strong interaction (QCD). In particular bound states and resonances constituting of $W$-s (W-hadrons) may appear.
We have already discussed a possibility to interpret the would-be CDF $Wjj$ excess~\cite{CDF} in terms of such state~\cite{AZ12}.
\includegraphics{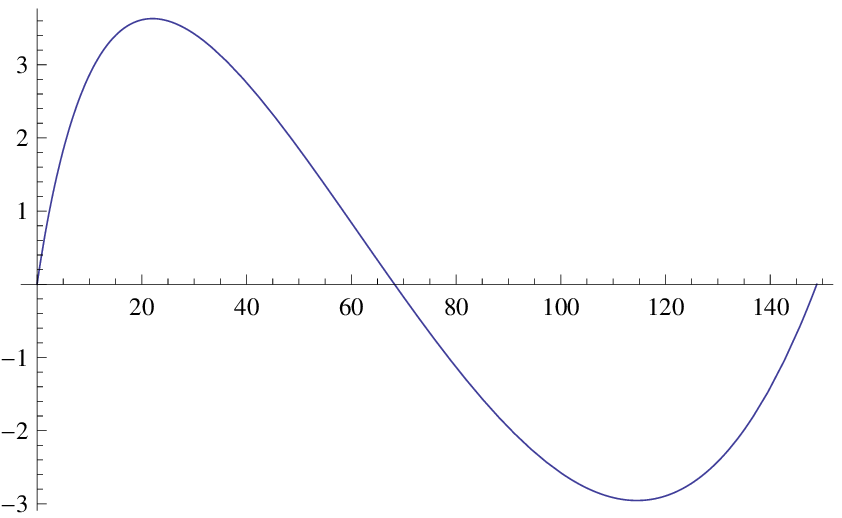}
\begin{flushleft}
{\small
Fig. 1. Behavior of the effective coupling $g_{eff}(t),\,t=G\,p^2$; $g_{eff}(t)\,=\,0 $ for $t\,>\,148$ .}
\end{flushleft}

\section{Scalar bound state of two W-s}

In the present work we apply these considerations along with some results of work~\cite{AZ11} to data indicating on a possible excess in  $\gamma\,\gamma$ and $l^+\,\l^+\,l^-\,l^-$ production at LHC~\cite{LHC1,LHC2} in region of invariant mass $120\,-\,130\,GeV$.

Let us assume that this excess is due to existence of
bound state $X$ of two $W$ with mass $M_s$. This state $X$ is assumed to have
spin 0 and weak isotopic spin also 0. Then vertex of $XWW$ interaction has the following form
\begin{equation}
\frac{G_X}{2}\,\,W_{\mu \nu}^a\,W_{\mu \nu}^a\,X\,\Psi\,;
\label{XWW}
\end{equation}
where $\Psi$ is a Bethe-Salpeter wave function of the bound state. Again due to gauge invariance there is also three-boson term
\begin{equation}
-\,g\,G_X\,\epsilon_{abc}\,W_{0\,\mu \nu}^a\,W_{\mu}^b\,W_{\nu}^c\,X\,;\label{XWWW}
\end{equation}
and four-boson term also. In what follows we use expressions~(\ref{XWW}, \ref{XWWW}).
The main interactions forming the bound state are just non-perturbative interactions~(\ref{FFF}, \ref{XWW}). This means that we take into account exchange of vector boson $W$ as well as of scalar bound state $X$ itself. In diagram form the corresponding Bethe-Salpeter equation is presented in Fig. 2.
We expand the kernel of the equation in powers of $M_W^2$ and $M_X^2$ and obtain the following equation with introduction of more
suitable variable
$$
z\,=\,\frac{G^2 (p^2)^2}{64\,\pi^2}\,;\quad t\,=\,\frac{G^2 (p^2)^2}{64\,\pi^2}\,;
$$
where~$p$~is~external~momentum~and~$q$~is the~integration momentum.
\begin{widetext}
\begin{eqnarray}
& &\Psi_0(z) = 4 \int_0^{z'_0}\Psi_0(t) dt - \frac{2}{3 z}\int_0^z \Psi_0(t) t dt+\frac{4}{3 \sqrt{z}}\int_0^z \Psi_0(t)\sqrt{t}\,dt +\frac{4 \sqrt{z}}{3}\int_z^{z'_0} \frac{\Psi_0(t)}{\sqrt{t}} dt -\frac{2 z}{3}\int_z^{z'_0}\frac{\Psi_0(t)}{t} dt+\nonumber\\
& &\frac{g}{4\, \pi}\biggl(-\,\frac{1}{z}\int_0^{z} \Psi_0(t) \sqrt{t}\,dt+ \frac{3}{\sqrt{z}}\int_0^z \Psi_0(t) dt +3\int_z^{z'_0} \frac{\Psi_0(t)}{\sqrt{t}} dt - \sqrt{z}\int_z^{z'_0} \frac{\Psi_0(t)}{t} dt\biggr)+\mu \biggl(-\,\frac{1}{z}\int_0^{z} \Psi_0(t) \sqrt{t}\,dt+\nonumber\\
& &\frac{2}{\sqrt{z}}\int_0^z \Psi_0(t) dt +6\int_0^{z'_0} \frac{\Psi_0(t)}{\sqrt{t}} dt + 2 \sqrt{z}\int_z^{z'_0} \frac{\Psi_0(t)}{t} dt-z \int_z^{z'_0} \frac{\Psi_0(t)}{t \sqrt{t}} dt \biggr)-\mu_s \biggl(\frac{1}{8\,z\,\sqrt{z}}\int_0^{z} \Psi_0(t) t\,dt-\nonumber\\
& &\frac{25}{64\,z}\int_0^{z} \Psi_0(t) \sqrt{t}\,dt+ \frac{19}{64\,\sqrt{z}}\int_0^z \Psi_0(t) dt +\frac{11}{8}\int_0^{z} \frac{\Psi_0(t)}{\sqrt{t}} dt +\frac{19}{16}\int_z^{z'_0} \frac{\Psi_0(t)}{\sqrt{t}} dt +\frac{5 \sqrt{z}}{16} \int_z^{z'_0} \frac{\Psi_0(t)}{t} dt-\label{BSX}\\
& &\frac{5 z}{64 \sqrt{z}}\int_z^{z'_0} \frac{\Psi_0(t)}{t \sqrt{t}}dt-\frac{z \sqrt{z}}{64 }\int_z^{z'_0} \frac{\Psi_0(t)}{t^2} dt\biggr)-\frac{\kappa}{12\,\pi}\biggl(\frac{1}{2 z }\int_0^z \Psi_0(t)\sqrt{t}\, dt+\frac{3}{2 \sqrt{z}}\int_0^z \Psi_0(t) dt +\frac{3}{2 }\int_z^{z'_0} \frac{\Psi_0(t)}{\sqrt{t}} dt+\nonumber\\
& &\frac{\sqrt{z}}{2 }\int_z^{z'_0} \frac{\Psi_0(t)}{t} dt\biggr)\,;\quad \mu\,=\,\frac{G\,M_W^2}{6\,\pi}\,;\quad \mu_s\,=\,\frac{G\,M_s^2}{6\,\pi}\,;\quad \kappa\,=\,\frac{G_X^2}{G}\,.\nonumber
\end{eqnarray}
\end{widetext}
Gauge electro-weak coupling $g$ enters due to diagrams of the second line of Fig. 2. Upper limit $z'_0$ is introduced for the sake of generality due the experience of works~\cite{BAA04} - \cite{AZ11}, according to which $z'_0$ may be either $\infty$ or some finite quantity. That is $z'_0$ is defined in a process of solving an equation. From the physical point of view
an effective cut-off $z'_0$ bounds a "low-momentum" region where our non-perturbative effects act and we consider the equation at interval $[0,\, z'_0]$ under condition
For form-factor of interaction~(\ref{FFF}) the upper limit $ z_0$~(\ref{z0}) is defined in work~\cite{AZ11}.
\begin{picture}(230,140)
{\thicklines
\put(35,100.5){\line(-3,2){30}}
\put(35,100.5){\line(-3,-2){30}}
\put(35,100.5){\circle*{6}}
\put(35,101.5){\line(1,0){20}}
\put(35,99.5){\line(1,0){20}}}
\put(18,119.5){p}
\put(18,80.5){-p}
\put(68.5,100){=}
{\thicklines
\put(112.5,100.5){\line(-3,2){30}}
\put(112.5,100.5){\line(-3,-2){30}}
\put(86.5,121.5){p}
\put(84.5,76.5){-p}
\put(112.5,100.5){\circle*{6}}
\put(95,89){\line(0,1){20.8}}
\put(95,89){\circle{5}}
\put(95,112){\circle{5}}
\put(112.5,101.5){\line(1,0){20}}
\put(112.5,99.5){\line(1,0){20}}}
\put(145,100){+}
{\thicklines
\put(187.5,100.5){\line(-3,2){30}}
\put(187.5,100.5){\line(-3,-2){30}}
\put(161.5,121.5){p}
\put(159.5,76.5){-p}
\put(187.5,100.5){\circle*{6}}
\put(169,89){\line(0,1){21}}
\put(171,89){\line(0,1){21}}
\put(170,89){\circle{7}}
\put(170,89){\circle{3}}
\put(170,112){\circle{7}}
\put(170,112){\circle{3}}
\put(187.5,101.5){\line(1,0){20}}
\put(187.5,99.5){\line(1,0){20}}
\put(215,100){+}
{\thicklines
\put(102.5,40.5){\line(-3,2){30}}
\put(102.5,40.5){\line(-3,-2){30}}
\put(76.5,60.5){p}
\put(74.5,16.5){-p}
\put(102.5,40.5){\circle*{6}}
\put(85,29){\line(0,1){21}}
\put(85,29){\circle*{1}}
\put(85,52){\circle{5}}}
\put(102.5,41.5){\line(1,0){20}}
\put(102.5,39.5){\line(1,0){20}}}
\put(140,40){+}
{\thicklines
\put(192.5,40.5){\line(-3,2){30}}
\put(192.5,40.5){\line(-3,-2){30}}
\put(166.5,60.5){p}
\put(164.5,16.5){-p}
\put(192.5,40.5){\circle*{6}}
\put(175,29){\line(0,1){23}}
\put(175,29.2){\circle{5}}
\put(175,45){\circle*{1}}
\put(192.5,41.5){\line(1,0){20}}
\put(192.5,39.5){\line(1,0){20}}}
\end{picture}

\begin{flushleft}
{\small
Fig. 2. Diagram representation of Bethe-Salpeter
equation for W-W bound state. Black spot corresponds to BS wave function. Empty circles correspond to point-like anomalous
three-gluon vertex~(\ref{FFF}), double circle -- XWW vertex~(\ref{XWW}). Simple point -- usual gauge triple $W$ interaction. Double line -- the bound state $X$, simple line -- W. All momenta are zero.}
\end{flushleft}

Bethe Salpeter wave function is normalized by condition $\Psi_0(0)\,=\,1$, which corresponds to the following condition
\begin{eqnarray}
& &4 \int_0^{z'_0}\Psi_0(t) dt\,+\,\frac{2 \sqrt{2}}{\pi}\int_0^{z_0} \frac{g\,F(t)}{\sqrt{t}}\, dt\,+\nonumber\\
& &\frac{3}{32\,\pi^2}\int_\mu^{z'_0} \frac{g^2\,\Psi_0(t)}{t}\, dt\,=\,1\,.\label{cond}
\end{eqnarray}
In diagram form this condition is presented at Fig. 3.
\begin{picture}(230,110)
{\thicklines
\put(30,80){$\textbf{1} \quad \textbf{=}$}
{\thicklines
\put(92.5,80.5){\line(-3,2){30}}
\put(92.5,80.5){\line(-3,-2){30}}
\put(92.5,80.5){\circle*{6}}
\put(75,69){\line(0,1){20.8}}
\put(75,69){\circle{5}}
\put(75,92){\circle{5}}}
\put(92.5,81.5){\line(1,0){20}}
\put(92.5,79.5){\line(1,0){20}}}
\put(135,80){+}
{\thicklines
\put(182.5,80.5){\line(-3,2){30}}
\put(182.5,80.5){\line(-3,-2){30}}
\put(182.5,80.5){\circle*{3}}
\put(182.5,80.5){\line(-1,0){17}}
\put(165,69){\line(0,1){11.8}}
\put(165,69){\circle{5}}
\put(182.5,81.5){\line(1,0){20}}
\put(182.5,79.5){\line(1,0){20}}
\put(220,80){+}
{\thicklines
\put(92.5,20.5){\line(-3,2){30}}
\put(92.5,20.5){\line(-3,-2){30}}
\put(92.5,20.5){\circle*{3}}
\put(92.5,20.5){\line(-1,0){18}}
\put(75,20.5){\line(0,1){13}}
\put(75,32){\circle{5}}}
\put(92.5,21.5){\line(1,0){20}}
\put(92.5,19){\line(1,0){20}}}
\put(135,20){+}
{\thicklines
\put(182.5,20.5){\line(-3,2){30}}
\put(182.5,20.5){\line(-3,-2){30}}
\put(182.5,20.5){\circle*{6}}
\put(165,9){\line(0,1){23}}
\put(165,9.2){\circle*{1}}
\put(165,25){\circle*{1}}
\put(182.5,21.5){\line(1,0){20}}
\put(182.5,19){\line(1,0){20}}}
\end{picture}
\begin{flushleft}
{\small 
Fig. 3. Diagram representation of normalization condition of Bethe-Salpeter  wave function.
Four leg vertex corresponds to interaction~(\ref{XWWW}). Other notations are the same as at Fig.~2.}
\end{flushleft}

We shall solve equation~(\ref{BSX}) by iterations.
Let us formulate the first approximation to equation~(\ref{BSX}). The first five terms of rhs of~(\ref{BSX}) will present the simplest zero approximation. We have in addition normalization condition~(\ref{cond}). There are few solutions of this set of equations but only one of them leads to positive $M_X^2$. It reads
\begin{eqnarray}
& &\Psi_1(z)=\frac{\pi}{2}\,G^{21}_{15}\bigl(z|^0_{1,0,1/2,-1/2,-1}\bigr)\,+\nonumber\\
& &C_1\,G^{20}_{0}\bigl(z|_{1,1/2,-1/2,-1}\bigr)\,+\label{Psi1}\\
& &C_2\,G^{10}_{0}\bigl(-\,z|_{1,1/2,-1/2,-1}\bigr); \;z'_0\,=\,44.151234 ;\nonumber\\
& &C_1\,=\,3.05437 ;\quad C_2\,=\,-0.0011964\,.\nonumber
\end{eqnarray}
where we again use Meijer functions~\cite{BE}.
Now we use solution~(\ref{Psi1}) and obtain parameter $\kappa$~(\ref{BSX}) with the aid of normalization condition for
$XWW$ coupling~(\ref{XWW}). In diagram form the condition is presented at Fig. 4. 

Namely we have
\begin{eqnarray}
& &\frac{\kappa}{8\,\pi}\biggl(9\,I_0\,-\,\frac{25}{16\,\pi}\,D^2\biggr)\,=\,1\,;\nonumber\\
& &I_0 = \int_0^{z'_0}\frac{\Psi_1^2(z)\,dz}{\sqrt{z}} ;\; D = \int_0^{z'_0}\frac{\sqrt{g}\,\Psi_1(z)\,dz}{\sqrt{z}} .\label{Psi0}
\end{eqnarray}
\begin{picture}(200,35)
{\thicklines
\put(20,10){$\textbf{1} \quad \textbf{=}$}
\put(55,10){\line(1,0){15}} \put(55,11){\line(1,0){15}}
\put(95,10){\line(1,0){15}}
\put(82,11){\oval(25,15)}
\put(95,11){\line(1,0){15}}
\put(70,10.5){\circle*{5}}
\put(95,10.5){\circle*{5}}
\put(150,10.5){\circle*{5}}
\put(175,10.5){\circle*{3}}
\put(118,10){$\textbf{+}$}
\put(135,10){\line(1,0){15}} \put(135,11){\line(1,0){15}}
\put(200,10){\line(1,0){15}}
\put(162,11){\oval(25,15)}
\put(187,11){\oval(25,15)}
\put(200,11){\line(1,0){15}}}
\put(200,10.5){\circle*{5}}
\end{picture}
\begin{flushleft} 
{\small
Fig. 4. Diagrams for normalization condition of $X\,W\,W$-vertex. Four-leg vertex corresponds to vertex~(\ref{FFAA})  being proportional to $gG$. }
\end{flushleft}
With $\Psi_1$~(\ref{Psi1}) we obtain from~(\ref{Psi0})
\begin{equation}
\kappa\,=\,0.592411\,.\label{kappa}
\end{equation}
Then we multiply equation~(\ref{BSX}) by $\Psi_{1}(z)$ from the right and integrate the result by $z$ in interval $(0,\,z'_0)$. It is easy to see by changing the order in double integrals, that all terms being of zero order vanish, and we
have the following equation
\begin{eqnarray}
& &-\,\mu_s \biggl(\frac{3\,I_1}{64}-\frac{5\,I_2}{64}+\frac{95\,I_3}{64}+\frac{11\,I_4}{8}-\frac{I_5}{64}\biggr)\,+\nonumber\\
& &\mu\bigl(-I_1+3\,I_2+14\,I_3+6\,I_4\bigr)-\label{pert}\\
& &\frac{\kappa}{12\,\pi}\bigl(I_2+3\,I_3\bigr)\,+\frac{3\,I_{g3}-I_{g2}}{4\,\pi};\nonumber
\end{eqnarray}
where
\begin{eqnarray}
& &I_1\,=\,\int_0^{z'_0}\frac{\Psi_1(z)\,dz}{z\sqrt{z}}\,\int_0^z\Psi_1(t) t\,dt\,;\nonumber\\
& & I_2\,=\,\int_0^{z'_0}\frac{\Psi_1(z)\,dz}{z}\,\int_0^z\Psi_1(t) \sqrt{t}\,dt\,;\nonumber\\
& &I_3\,=\,\int_0^{z'_0}\frac{\Psi_1(z)\,dz}{\sqrt{z}}\,\int_0^z\Psi_1(t)\,dt\,;\nonumber\\
& &I_4\,=\,\int_0^{z'_0}\Psi_1(z)\,dz\,\int_0^z\frac{\Psi_1(t)\,dt}{\sqrt{t}}\,;\label{INT}\\
& & I_5\,=\,\int_0^{z'_0}\frac{\Psi_1(z)\,dz}{z^2}\,\int_0^z\Psi_1(t) t\sqrt{t}\,dt\,;\nonumber\\
& & I_{g2}\,=\,\int_0^{z'_0}\frac{g\,\Psi_1(z)\,dz}{z}\,\int_0^z\Psi_1(t) \sqrt{t}\,dt\,;  \nonumber\\
& &I_{g3}\,=\,\int_0^{z'_0}\frac{g\,\Psi_1(z)\,dz}{\sqrt{z}}\,\int_0^z\Psi_1(t)\,dt\,.\nonumber
\end{eqnarray}
Now we define running coupling $g$
\large{
\begin{equation}
g\,=\,\frac{g(M_W)}{\sqrt{1+\frac{5\,g^2(M_W)}{24\,\pi^2}\ln\biggl(1+\frac{8\,\pi\sqrt{z}}{G\,M_W^2}\biggr)}}\,.
\label{rung}
\end{equation}}
It enters in integrals~(\ref{kappa}, \ref{INT}).
We introduce $M_X\,=\,125\,GeV$ that means
\begin{equation}
\mu_s\,=\,\mu\,\frac{125^2}{80.4^2}\,; \label{rung}
\end{equation}
and perform necessary calculations. So we choose a solution with mean value $M_X = 125\,GeV$ of the ATLAS and the CMS results~\cite{LHC1,LHC2}, then we have unique solution with the following parameters
\begin{equation}
 G_X = 0.000666\,GeV^{-1} ;\;
 G = \frac{0.00484}{M_W^2}\,.\label{GGX}
\end{equation}
Result~(\ref{GGX}) means parameter of anomalous triple interaction~(\ref{FFF}) with account of relation~(\ref{Glam})
\begin{equation}
 \lambda\,=\,-\frac{G\,M_W^2}{g(0)}\,=\,-\,0.00744\,;\label{lambda}
\end{equation}
which doubtless agrees limitations~(\ref{lambda1}).\\

\section{Comparison to experiments}
Thus we have scalar state $X$ with coupling~(\ref{XWW}, \ref{GGX}). In calculations of decay parameters and cross-sections we use CompHEP package~\cite{Boos}.
Cross-section of $X$ production at LHC with $\sqrt{s}\,=\,7\,TeV$ reads
\begin{equation}
\sigma_X\,=\,\sigma(p+p\to X+...)\,=\,0.184\,pb \,\label{Section}
\end{equation}
Parameters of $X$-decay are the following
\begin{eqnarray}
& &\Gamma_t(X) = 0.000502\,GeV ;\nonumber\\
& & BR(X \to \gamma \gamma)\,=\,0.430 ;\nonumber\\
& & BR(X \to \gamma Z)\,=\,0.305 ;\nonumber\\
& &BR(X \to 4\,l (\mu, e))\,=\,0.000577 ;\nonumber\\
& &BR(X \to b\, \bar b)\,=\,  0.000024\,.\nonumber\\ 
& & BR(X \to \gamma e^+ e^-)\,=\,0.0231 ;\label{decay}\\
& & BR(X \to \gamma \mu^+ \mu^-)\,=\,0.016 ;\nonumber\\
& & BR(X \to \gamma \tau^+ \tau^-)\,=\,0.0125 ;\nonumber\\
& &BR(X \to \gamma u \bar u)\,=\,0.0478 ;\nonumber\\
& & BR(X \to  \gamma c \bar c)\,=\,0.0368 ;\nonumber\\
& & BR(X \to \gamma d \bar d)\,=\,0.0446 ;\nonumber\\
& &BR(X \to \gamma s \bar s))\,=\,0.0430 ;\nonumber\\
& &BR(X \to \gamma b \bar b)\,=\,  0.0416\,.\nonumber  
\end{eqnarray}
For decay $X \to b \bar b$ we calculate the evident triangle diagram and use $m_b(125\,GeV) \simeq 2.9\,GeV$. Branching ratios for decays to 
other fermion pairs are even smaller.

Experimental data give in the region of the would-be state the following limitations for $\sigma_{\gamma \gamma}\,=\,\sigma_X\,BR(X \to \gamma \gamma)$ 
\begin{eqnarray}
& &\sigma_{\gamma \gamma}\,<\,3.8\,\sigma(SM)\,;\quad \label{limit}\\
& &\sigma_{\gamma \gamma}\,<\,3.6\,\sigma(SM)\,;\quad \sigma_{\gamma \gamma}\,<\,0.135\,pb\,.\nonumber
\end{eqnarray}
Here $\sigma(SM)\,\simeq\,0.04\,pb$ is the Standard Model value for the quantity under discussion, upper line correspond to ATLAS data~\cite{ATLAS1} and the lower line correspond to CMS data~\cite{CMS1}. Firstly both limitations are quite consistent. Secondly our value for the same quantity from~(\ref{Section}, \ref{decay}) reads 
\begin{equation}
\sigma_{\gamma \gamma}\,=\,0.077\,pb\,;\label{CS0}
\end{equation}
that also agrees limitations~(\ref{limit}), however it essentially exceeds the SM value $\sigma(SM)$. At this point it is advisable to discuss accuracy of our approximations. 
The former experience concerning both applications to Nambu -- Jona-Lasinio model in QCD~\cite{BAA06,AVZ06,AVZ09} and to the electro-weak interaction~\cite{BAA09,AZ11} shows that average accuracy of the method is around 10\% in values of different parameters. So we may assume, that in the present estimations of coupling constant $G_X$ we also have the same accuracy. For the cross-section this means possible deviation up to 20\% of the calculated value. Thus we would change~(\ref{CS0}) to  
the following result
\begin{equation}
\sigma_{\gamma \gamma}\,=\,(0.077\pm 0.015)\,pb\,;\label{CS1}
\end{equation}
Branching ratios~(\ref{decay}) do not depend on the value of $G_X$, so we assume their accuracy being considerably better than in~(\ref{CS1}).
In any case result~(\ref{CS1}) agrees~(\ref{limit}).

There are also indications for some excess around $125\,GeV$ in four leptons states.
With our numbers~(\ref{Section}, \ref{decay}) we have for decay $X \to l^+\,l^+\,l^-\,l^- (l = \mu, e)$: $\sigma \times BR\,=\,(0.00011\pm 0.00002)\,pb$. For integral luminosity $L\,=\,4.8\,10^3\,pb^{-1}$~\cite{ATLAS1, CMS1} we have for number of events
\begin{equation}
N(4\,l)\,=\,\sigma\times BR\times L\,=\,(0.51\pm 0.10) ev\,;\label{Fl}
\end{equation}
i.e. close to one event. This result also essentially exceeds the SM expectations. As a matter of fact ATLAS~\cite{ATLAS2}  has three events and CMS~\cite{CMS2} -- two ones in the region under consideration with estimated background rather smaller than one event. In any case our estimation~(\ref{Fl}) has no contradiction with data as well as the usual SM Higgs boson interpretation. In the future more precise experiments  at LHC the essential distinctions of our scheme and the SM Higgs boson variant could manifest themselves and decisively discriminate different variants. The distinctions refer to 
$\sigma_{\gamma \gamma}$~(\ref{CS1}) and also to the four-lepton channel~(\ref{Fl}). 

We would emphasize importance of channel $X \to \gamma\, l^+ l^-$. For this decay mode from~(\ref{Section}, \ref{decay}) we predict
\begin{equation}
\sigma_X\,BR(X \to \gamma l^+ l^-)\,=\,(0.0073 \pm 15)\,pb;
\label{All}
\end{equation}
that gives $N = 35 \pm 7$ events for already achieved luminosity~\cite{LHC1,LHC2}. This channel may serve for an accurate test of our results because the SM value for quantity~(\ref{All}) gives around 5 events~\cite{Gainer}.

There is one point in data~\cite{CMS1}, which provides hints against the SM option and, on the contrary, on behalf of our variant. There are data of~\cite{CMS1} dealing with two-jets tag, which singles channel of $X (H)$ production via vector boson fusion. We calculate the effect for this channel within our approach and obtain $\sigma_{VBF}\,=\,0.079\,pb$. Taking into account~(\ref{decay}) and efficiency in~\cite{CMS1} for such process $\simeq 0.037$ we obtain $6$ events of $\gamma\,\gamma$ decay of $X$, what by no means contradict CMS data (see~\cite{CMS1}, Fig. 1(b)). The estimate for the SM Higgs gives here less than one event. Of course, there is no contradiction yet, but nevertheless we may state a trend for better agreement with data of the present variant. 

The main difference of our predictions with the SM results consist in decay channel $X\to b \bar b$. For SM Higgs which is usually considered for explanation of would-be $125\,GeV$ state this decay is dominant, whereas our result~(\ref{decay}) gives extremely small $BR\,\simeq 3\,10^{-5}$. We would emphasize that SM Higgs interpretation could not be considered as proved unless $b \bar b$ channel with the proper intensity would be detected.  

\section{Conclusion}

Thus we have an alternative interpretation of LHC $125\,GeV$ phenomenon.
The overall data do not contradict both the SM Higgs option and the scalar W-hadron, which we discuss here. However our estimates of the effects are as a rule rather more than the SM Higgs predictions. It seems, that data favor just larger values. The forthcoming increasing of the integral luminosity will
undoubtedly discriminate this two options. In case of future result being in favor of scalar W-hadron, we need additional comparison of our predictions with results of other possibilities, e.g. fermiophobic variant~\cite{fph1,fph2}, which in application to data~\cite{LHC1,LHC2} is discussed in paper~\cite{fph3}.

The authors express gratitude to E.E. Boos and V.I. Savrin for valuable discussions.
The work was supported in part by grant of Russian Ministry of Education and Science NS-3920.2012.2.

\end{document}